# Anomalous doping evolution of superconductivity and quasiparticle interference in $Bi_2Sr_2Ca_2Cu_3O_{10+\delta}$ trilayer cuprate

Zhenqi Hao[1,*], Changwei Zou[1,*], Xiangyu Luo[2], Yu Ji[1], Miao Xu[1], Shusen Ye[1], Xingjiang Zhou[2], Chengtian Lin[3], and Yayu Wang[1,4†]

[1]*State Key Laboratory of Low Dimensional Quantum Physics, Department of Physics, Tsinghua University, Beijing 100084, P. R. China*

[2]*Beijing National Laboratory for Condensed Matter Physics, Institute of Physics, Chinese Academy of Sciences, Beijing 100190, P. R. China*

[3]*Max Planck Institute for Solid State Research, Heisenbergstr 1, D-70569 Stuttgart, Germany*

[4]*Frontier Science Center for Quantum Information, Beijing 100084, P. R. China*

[*]These authors contributed equally to this work.

[†]yayuwang@tsinghua.edu.cn

We use scanning tunneling microscopy to investigate $Bi_2Sr_2Ca_2Cu_3O_{10+\delta}$ trilayer cuprates from the optimally doped to overdoped regime. We find that the two distinct superconducting gaps from the inner and outer $CuO_2$ planes both decrease rapidly with doping, in sharp contrast to the nearly constant $T_C$. Spectroscopic imaging reveals the absence of quasiparticle interference in the antinodal region of overdoped samples, showing an opposite trend to that in single- and double-layer compounds. We propose that the existence of two types of inequivalent $CuO_2$ planes and the intricate interaction between them are responsible for these highly anomalous observations in trilayer cuprates.

Despite the lack of consensus regarding the mechanism of superconductivity in cuprates, there are two well-established trends about the critical temperature $T_C$. One is the dome-shaped doping dependence of $T_C$, which reaches a peak at hole density $p \sim 0.16$ [1]. The other is the variation of maximum $T_C$ with the number of $CuO_2$ planes per unit cell, which is the highest in the trilayer compound of a homologous series [2]. The origin of enhanced superconductivity in trilayer cuprate has attracted tremendous interest because it is not only a key issue regarding the mechanism, but may also provide valuable clues for finding even higher $T_C$ superconductors. Most theoretical models focus on the unique character of trilayer cuprates, namely the existence of inequivalent inner and outer $CuO_2$ planes. It has been proposed that the proximity between the underdoped inner plane (IP) with large pairing strength and the overdoped outer plane (OP) with strong phase stiffness provides an optimal condition for superconductivity [3,4]. The Josephson tunneling between Cooper pairs in the inequivalent $CuO_2$ planes within a unit cell has also been proposed to be crucial for the enhanced $T_C$ [5,6].

The $Bi_2Sr_2Ca_2Cu_3O_{10+\delta}$ (Bi-2223) system represents an ideal platform for studying the peculiar physics of trilayer cuprates because it can be easily cleaved for surface sensitive probes such as angle-resolved photoemission spectroscopy (ARPES) [7-13] and scanning tunneling microscopy (STM) [14-16]. The schematic crystal structure of Bi-2223 is depicted in Fig. 1 inset. The Cu atoms in the IP form a $CuO_4$ plaquette without apical oxygen, while the Cu atoms in the two OPs are surrounded by a $CuO_5$-pyramid with one apical oxygen. The different structural and chemical environments of the $CuO_2$ planes may have profound influences on their superconducting properties. ARPES experiments on Bi-2223 have revealed two sets of Fermi surfaces (FSs), in which the OP is more heavily doped than the IP [10]. Moreover, the hybridization between the IP and OP has been found to modify the Bogoliubov band dispersion and enhance the superconducting gap [12].

A highly anomalous observation regarding Bi-2223 is that upon reaching the optimal doing, $T_C$ remains almost a constant with further increase of hole density [17-20]. This trend is an obvious violation of the dome-shaped $T_C$ phase line that is deemed universal to all curpates. To explain the extraordinarily robust superconductivity in overdoped Bi-2223, it was hypothesized that the IP remains at the optimal doping level despite increased oxygen content [19], similar to the interface between underdoped and overdoped cuprates [21]. However, there is no experimental data so far on the IP gap size of overdoped Bi-2223, leaving this conjecture untested. In fact, the overdoped regime of Bi-2223 has been little explored by spectroscopic techniques, thus even the basic information regarding the electronic structure is still unknown.

In this letter, we use STM to study the evolution of electronic structure in Bi-2223 from the optimally doped to overdoped regime. We find that despite the nearly constant $T_C$, the two superconducting gaps from the IP and OP both decrease rapidly with doping. Spectroscopic imaging reveals the systematic increase of hole density, but there is an absence of quasiparticle interference (QPI) in the antinodal region of overdoped Bi-2223. We propose that the distinct doping evolution of the two types of $CuO_2$ planes and the intricate interaction between them are essential for understanding these anomalous properties in trilayer cuprates.

A series of Bi-2223 crystals were grown by the travelling solvent floating zone method and post-annealed in different oxygen pressure and temperature conditions to achieve overdoping. Details about sample growth and treatment have been published previously [22,23]. Figure 1a shows the temperature dependent susceptibility measured by SQUID magnetometer on the optimally doped sample (labelled as OPT) and three overdoped samples with progressively increased hole density (labelled as OD1, OD2, and OD3). Remarkably, the critical temperatures are $T_C$ = 110 K, 107 K, 109 K and 108 K for the OPT, OD1, OD2 and OD3 samples, which confirm the finding of weakly doping dependent $T_C$ in overdoped Bi-2223 [17-20]. All the STM experiments are performed at $T$ = 5 K, and the differential conductance ($dI/dV$) spectroscopy is obtained by standard lock-in method.

Figure 1b displays the topography of an OPT crystal, in which the Bi atoms and structural supermodulations of the exposed BiO surface can be clearly resolved. Topographic images of the three overdoped samples are shown in Supplementary Fig. S1. The spatially averaged $dI/dV$ spectra for the four samples are plotted in Fig. 1c. The average superconducting gap size, defined by the position of the coherence peak, is Δ = 61, 42, 33 and 22 meV, respectively. The decrease of Δ with increasing hole density is consistent with the expectation for overdoping, and agrees with that obtained by interlayer tunneling spectroscopy on overdoped Bi-2223 [19]. In the most overdoped OD3 sample, Δ is reduced to around 36% of that in the OPT sample, which is highly striking given the nearly identical $T_C$.

Although the spatially averaged $dI/dV$ in Bi-2223 is similar to that in other cuprates, a closer examination of each individual $dI/dV$ curve reveals a highly unique feature. As shown by the representative $dI/dV$ curves in Fig. 2a, there are two coherence peaks that are symmetric with respect to the Fermi energy, corresponding to two different gaps from the inequivalent IP and OP respectively. The energy of the two peaks can be extracted more accurately by performing a double-Lorentzian fit of the raw data (see Supplementary Fig. S2), and the gap values denoted as $\Delta_{IP}$ and $\Delta_{OP}$ are summarized in Fig. 2b. The distribution of each gap has an approximate Gaussian form, and the average values of the two gaps are well-separated. The two gap sizes in the OPT sample are close to that determined by Raman spectroscopy on similar Bi-2223 [24]. With increasing doping, both $\Delta_{IP}$ and $\Delta_{OP}$ decrease systematically and rapidly. Due to the rather broad distribution of $\Delta_{IP}$ and $\Delta_{OP}$, the two-gap feature is smeared out in the averaged spectra displayed in Fig. 1c, as demonstrated in our previous report [16].

We have performed spectroscopic imaging on each sample by taking d$I$/d$V$ curves on a dense grid, which enables us to extract the momentum-space information through the QPI patterns. The conductance maps $g(r, E) = dI/dV (r, E)$ at selected energies for each sample are displayed in supplementary Fig. S3. In order to eliminate the influences of setpoint and static charge modulations, a common practice in analyzing the QPI data is to plot the ratio map $z(r, E) = g(r, E)/g(r, -E)$, which can be justified by the coherent admixture of electron and hole components for Bogoliubov quasiparticles [25,26]. The $z$ maps for the four samples at a series of bias voltages are displayed in Fig. 3a-d, which clearly reveal the scattering interference patterns. With varied energies, the QPI pattern and wavelength change systematically according to the dispersion relation of the quasiparticle wavevector $q$. For example, the stripe-like pattern in the first column of each panel is contributed by the $q_7$ components that will be shown later. With increasing doping, the energy range exhibiting pronounced QPI patterns becomes smaller, in accordance with the reduction of superconducting gap size.

By applying Fourier transform (FT) analysis on the $z$ map, the periodical modulations in real space can be converted into bright spots in $q$ space. The last columns in Fig. 3 display the FT maps corresponding to the second column $z$ maps of the four samples, where the QPI patterns are most pronounced. The bright spots in the FT images generally follow the 'octet model' [27], which describes the scattering interference between the 'hot spots', or joint density of state maximum, at the equal-energy contours of Bogoliubov quasiparticles. Owing to the $d$-wave gap function, there are seven wavevectors originated from the banana-shaped equal-energy contours, as depicted in Fig. 4a. Based on the systematic energy evolution of $q$, the underlying FS can be constructed [28], as exemplified in supplementary Fig. S4 for the OPT sample. The open symbols in Fig. 4b represent the QPI-derived FS for the four samples, and the broken lines are numerical fits by assuming a circular hole pocket. The enlargement of hole-type FS area from the OPT to OD3 samples is consistent with the increase of hole density, and further confirms the effectiveness of progressive doping.

Despite the similarity of the QPI patterns with that in other well-studied cuprates [25,26,29-32], there is a highly unusual trend revealed by Fig. 4b. With increasing doping, the $q$-section with observable QPI becomes shorter and further away from the antiferromagnetic Brillouin zone boundary (marked by the diagonal black broken line). Or in another word, there is an apparent absence of QPI signal in the antinodal region of the FS in overdoped Bi-2223. This trend is opposite to that observed in single- and double-layer Bi-based cuprates [29,30], where the $q$-section keeps increasing in the strongly overdoped regime due to the elongation of Fermi arc. With the highest $T_C$ and largest superfluid density, Bi-2223 is expected to have the most coherent quasiparticles over the entire Brillouin zone. The absence of QPI in the antinodal region represents a significant departure from the octet model, and another unexpected feature of overdoped Bi-2223.

Our STM experiments reveal two highly anomalous behaviors in overdoped Bi-2223: the rapid decrease of both $\Delta_{IP}$ and $\Delta_{OP}$ in the presence of nearly constant $T_C$, and the suppression of QPI in the antinodal region. Both trends are absent in the single- and double-layer Bi-based cuprates, thus are unique to the trilayer compound. We believe the key physics lies in the most unique trait of trilayer cuprates, namely the two types of inequivalent $CuO_2$ planes. Due to the limited experimental and theoretical studies on overdoped Bi-2223, below we will only give some phenomenological explanations for these anomalies.

The questions regarding what determines the $T_C$ of a cuprate and the overall dome-shaped dependence have been core issues in high $T_C$ superconductivity since its discovery. The most influential viewpoint is that $T_C$ is determined by two more fundamental temperature scales [33], the phase stiffness temperature and mean-field pairing temperature. In the overdoped regime with high superfluid density, $T_C$ is expected to decrease as the superconducting gap size reduces. It can naturally explain the $T_C$ dome, but is apparently not obeyed here by Bi-2223. To reconcile such discrepancy, it has been hypothesized that the IP of overdoped Bi-2223 remains at the optimal hole density despite increasing oxygen content, which helps maintain the highest $T_C$ [19]. However, the spatially resolved tunneling spectra shown in Fig. 2 directly demonstrate the rapid decrease of both $\Delta_{IP}$ and $\Delta_{OP}$, thus can safely rule out this possibility.

We find that a more applicable model for the anomalously high $T_C$ in overdoped Bi-2223 is by considering the effective superconducting gap size $\Delta_{SC}$ of the two inequivalent $CuO_2$ planes. It was proposed that due to the $d$-wave pairing symmetry in cuprates, $T_C$ is determined by the "coherent gap size" at the tip of Fermi arc because beyond that the coherent Bogoliubov quasiparticles are suppressed by the pseudogap [11,34,35]. Based on ARPES results, it was revealed that the OP and IP of Bi-2223 have different Fermi arc lengths and gap sizes. The $T_C$ of optimally doped sample is determined by the effective gap size $\Delta_{SC} = 21\pm3$ meV at the Fermi arc tip of the OP [11], as schematically illustrated in Fig. 4c, because that of the much shorter IP arc is not sufficient to support a $T_C \sim 110$ K. With increasing doping, the Fermi arc becomes longer but the $d$-wave gap size becomes smaller for both $CuO_2$ planes. In the heavily overdoped OD3 sample, $\Delta_{OP}$ shown in Fig. 2b is reduced to merely 21 meV. It has been shown that the energy gap detected by STM mainly reflects the gap size at the antinodal point, which has the maximum value in the $d$-wave gap function [36], so $\Delta_{SC}$ at the Fermi arc tip should be even smaller. On the other hand, $\Delta_{IP}$ is shown to be 35 meV in Fig. 2b, thus $\Delta_{SC}$ can reach 21 meV given the elongated Fermi arc (illustrated in Fig. 4d). This makes it possible to maintain a $T_C \sim 110$ K because it is well-known that for superconductors with multiple gaps, $T_C$ is determined by the largest gap size [12,37,38]. Therefore, the two inequivalent $CuO_2$ planes with complementary superconducting properties can give rise to a weakly doping dependent $T_C$ in trilayer cuprates. Note that for simplicity, the effect of interlayer hybridization on $\Delta_{IP}$ and $\Delta_{OP}$ [12] is not presented in the schematic gap function in Fig. 4c and 4d.

The absence of QPI in the antinodal region of overdoped Bi-2223 is another manifestation of the peculiar electronic structure of trilayer cuprates. In overdoped single- and double-layer cuprates, the elongation of Fermi arc has been shown to generate strong QPI features over an extended section in $k$-space [29,30,39]. In trilayer Bi-2223, however, the hybridization between the IP and OP strongly modifies the electronic structure. As revealed by ARPES band mapping on optimally doped Bi-2223 [12], the hybridization makes the dispersion near the antinodal region flatter. Such tendency is expected to become more pronounced with increasing doping, when the band dispersion of the heavily overdoped OP approaches the flat saddle point. As depicted schematically in Fig. 4d for the OD3 sample, the Bogoliubov quasiparticles of the OP are distributed over a much broader $k$-region in the antinodal direction. Consequently, the equal-energy contour no longer has a banana-shape, and there is no singular wavevector connecting the sharp "hot spots", which is required for the formation of well-defined QPI patterns.

In summary, spectroscopic imaging STM experiments on the overdoped regime of Bi-2223 reveal highly anomalous doping evolution of superconductivity and QPI patterns that are distinctively different from single- and double-layer compounds. We propose that the existence of two types of inequivalent $CuO_2$ planes and the intricate interaction between them are the underlying reasons for these anomalies. These results highlight the unique characteristics of trilayer cuprates, which are the key factors for achieving the highest $T_C$.

**Acknowledgements:** This work was supported by the NSFC grant No. 11534007, MOST of China grant No. 2017YFA0302900 and No. 2016YFA0300300, the Basic Science Center Project of NSFC under grant No. 51788104, NSFC Grant No. 11888101, and the Strategic Priority Research Program (B) of the Chinese Academy of Sciences (XDB25000000). This work is supported in part by the Beijing Advanced Innovation Center for Future Chip (ICFC).

**Figure Captions:**

FIG. 1. (a) Temperature dependent magnetic susceptibility of four Bi-2223 samples with different dopings, which exhibit nearly constant $T_C$. Inset: the crystal structure of Bi-2223 with inequivalent IP and OPs. (b) The topographic image of an OPT sample taken with tunneling current $I$ = 5 pA and bias voltage $V$ = -250 mV. (c) Spatially averaged $dI/dV$ curves of the OPT, OD1, OD2 and OD3 samples. The average superconducting gap size decreases significantly with increasing doping.

FIG. 2. (a) Representative individual $dI/dV$ spectra in the four samples showing two coherence peaks corresponding to the superconducting gaps from the IP and OP (denoted by $\Delta_{IP}$ and $\Delta_{OP}$), respectively. (b) In each sample the distribution of $\Delta_{IP}$ and $\Delta_{OP}$ can be fitted well by a Gaussian line shape. Both $\Delta_{IP}$ and $\Delta_{OP}$ decrease rapidly with increasing doping.

FIG. 3. (a) to (d): The conductance ratio maps $z(r, E) = g(r, E)/g(r, -E)$ at selected energies for the OPT, OD1, OD2, and OD3 samples, where the dispersive QPI patterns can be clearly visualized. The scale bar corresponds to 10 nm in each image. Note the systematic decrease of energy scales with increasing doping, which reflects the reduction of superconducting gap size. The last columns in each panel display the FT maps corresponding to the second column $z$ maps with most pronounced QPI patterns.

FIG. 4. (a) Schematic illustration of the $d$-wave gap structure and the QPI wavevectors $q_1 \sim q_7$ in the 'octet' model. (b) The underlying FSs of the four samples derived from their FT-QPI patterns, which exhibit systematic doping dependence. (c) and (d) Left panels: the schematic band structures of the IP (red) and OP (blue) for the OPT and OD3 samples. Right panels: the $k$-dependence of $\Delta_{IP}$ and $\Delta_{OP}$ with $d$-wave gap function. The solid blue and red dots indicate the tip of the Fermi arc with effective superconducting gap $\Delta_{SC} \sim 21$ meV for the two samples.

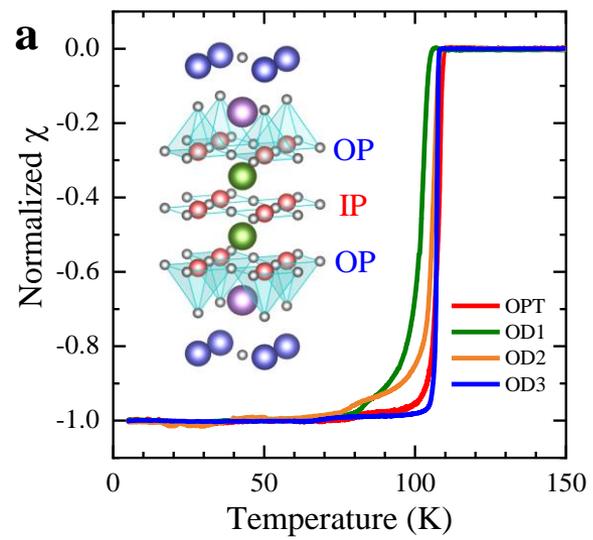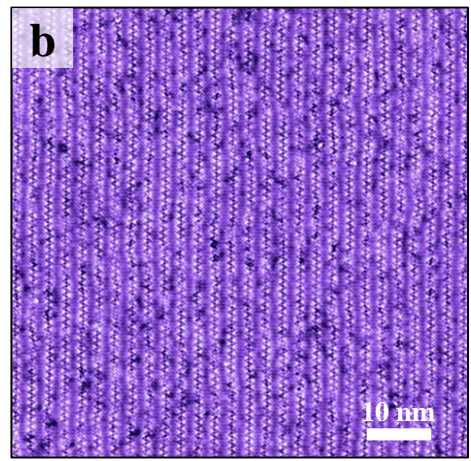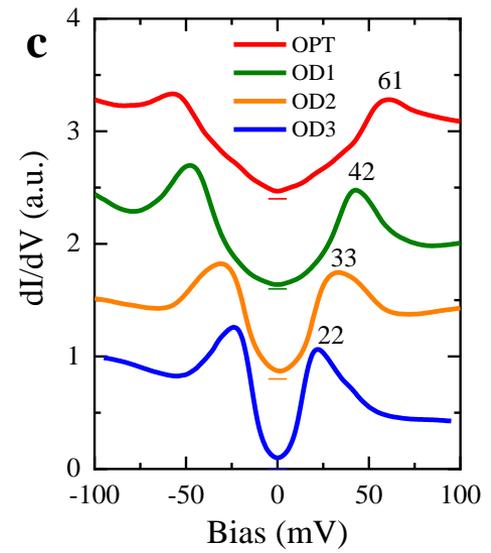

Figure 1

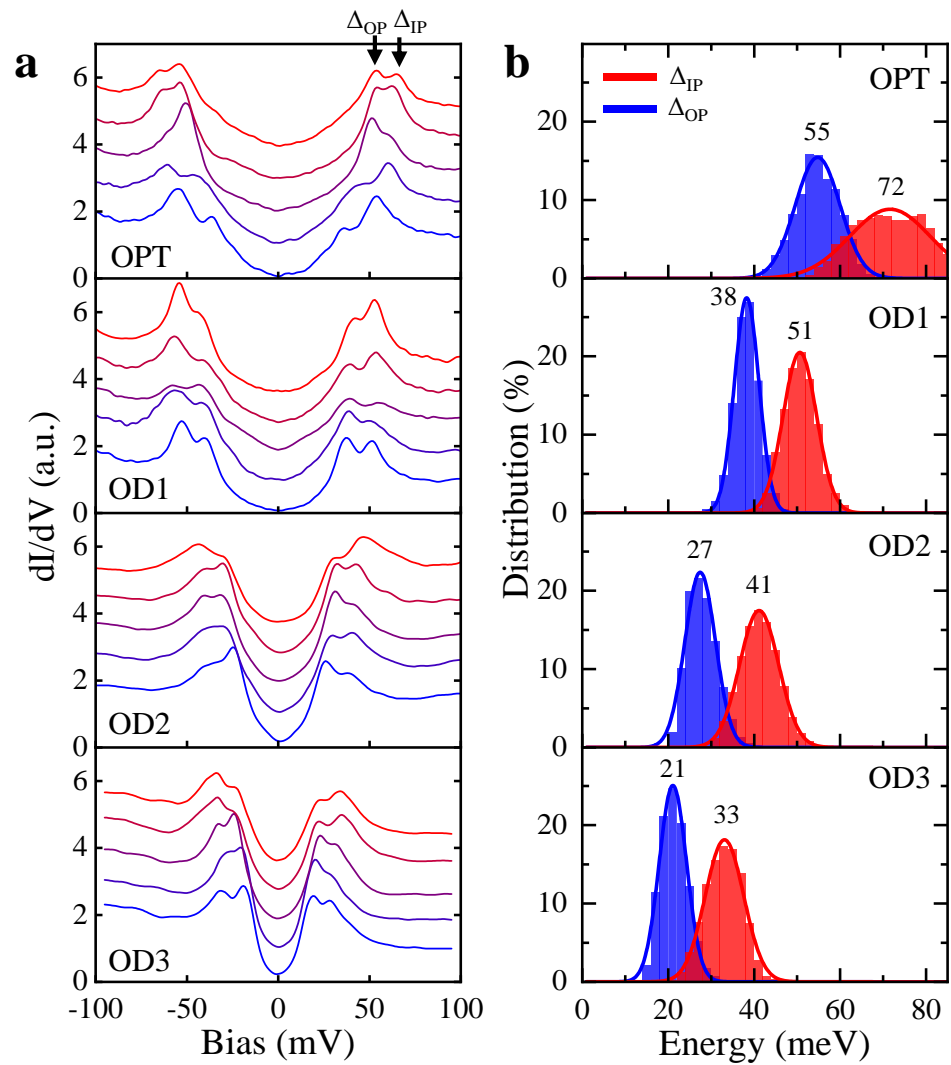

Figure 2

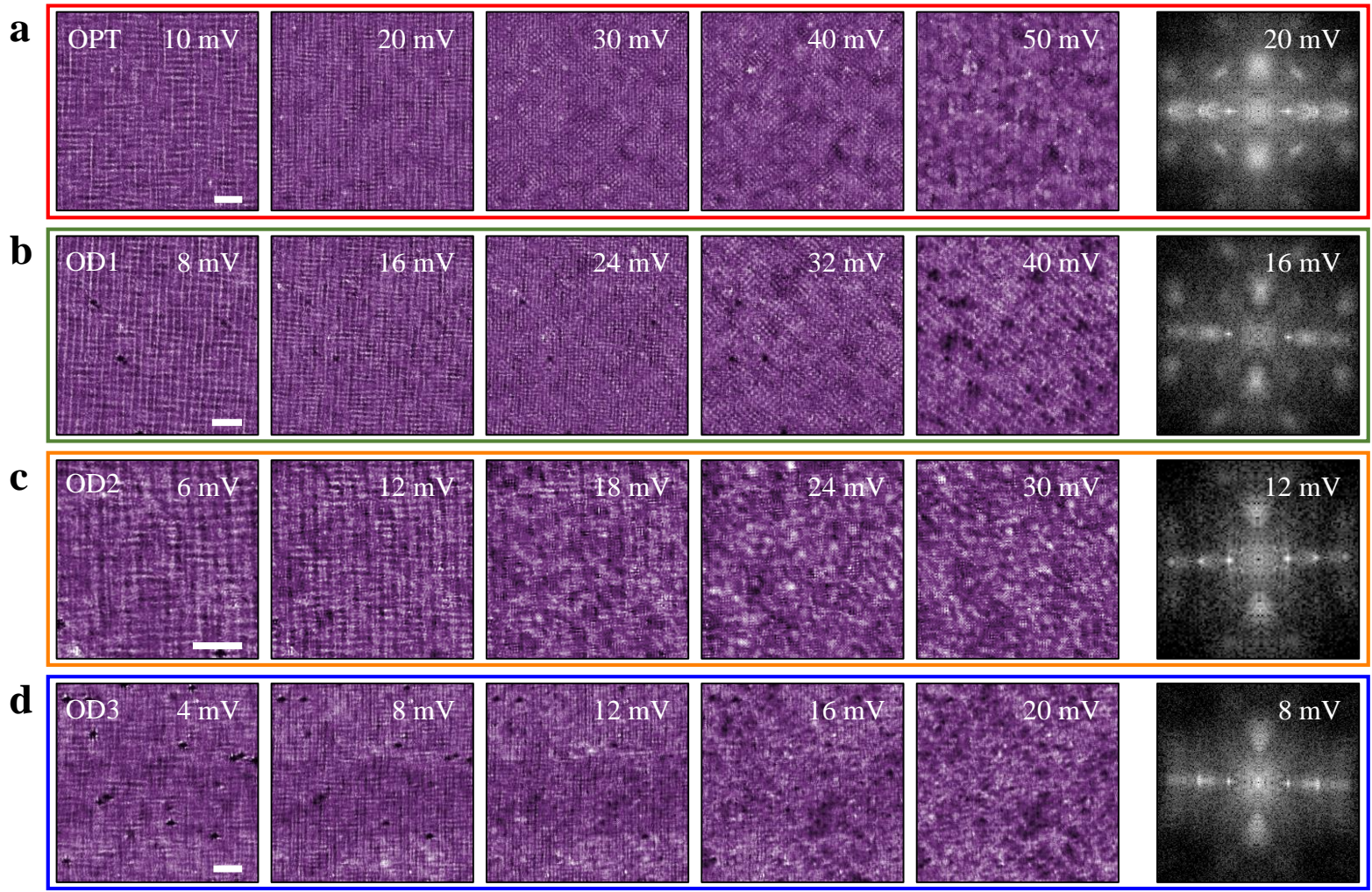

Figure 3

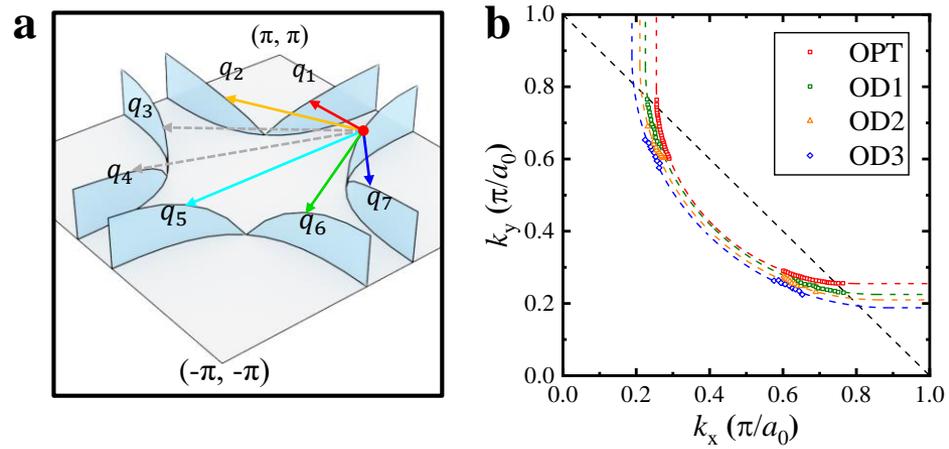
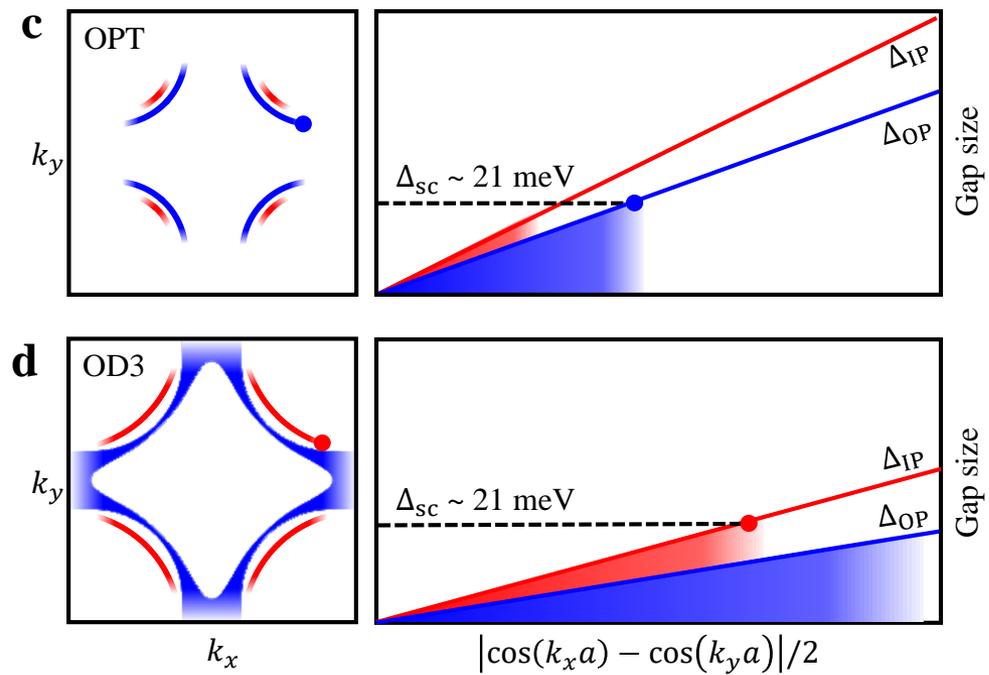

Figure 4